\documentclass[amsmath,amssymb,reprint,twocolumn,superscriptaddress, prb]{revtex4-1}

\usepackage{amsmath,amssymb,amsfonts}
\usepackage{amstext, mathrsfs, textcomp}
\usepackage{mathptmx}
\usepackage{bigints}  
\usepackage{nicefrac}

\usepackage{multirow}
\usepackage{dcolumn}
\usepackage{bm}

\usepackage{graphicx}
\usepackage{xcolor}
\usepackage[export]{adjustbox}

\usepackage[colorlinks=true, citecolor=blue, urlcolor=blue, linkcolor=blue]{hyperref}

\usepackage{changes}

\definechangesauthor[color=blue]{OJ}

\usepackage{listings}

\definecolor{codegreen}{rgb}{0,0.6,0}
\definecolor{codegray}{rgb}{0.5,0.5,0.5}
\definecolor{codepurple}{rgb}{0.58,0,0.82}
\definecolor{backcolour}{rgb}{0.95,0.95,0.92}

\lstdefinestyle{mystyle}{
backgroundcolor=\color{backcolour},
commentstyle=\color{codegreen},
keywordstyle=\color{blue},
numberstyle=\tiny\color{codegray},
stringstyle=\color{codepurple},
basicstyle=\ttfamily\scriptsize,
breakatwhitespace=false,
breaklines=true,
captionpos=b,
keepspaces=true,
numbers=left,
numbersep=5pt,
showspaces=false,
showstringspaces=false,
showtabs=false,
tabsize=2
}
\graphicspath{{Figures/}}

\newcommand{\TITLE}{PyMieDiff: A differentiable Mie scattering library}

\newcommand{\code}[1]{\texttt{#1}}

\begin{document}
\title{\TITLE}

\author{\firstname{Oscar K. C.} \surname{Jackson}}
\affiliation{Physics and Astronomy, Faculty of Engineering and Physical Sciences, University of Southampton, SO17 1BJ Southampton, UK}

\author{\firstname{Simone} \surname{De Liberato}}
\affiliation{Physics and Astronomy, Faculty of Engineering and Physical Sciences, University of Southampton, SO17 1BJ Southampton, UK}
\affiliation{Istituto di Fotonica e Nanotecnologie -- Consiglio Nazionale delle Ricerche (CNR), Piazza Leonardo da Vinci 32, Milano, Italy}
\affiliation{Sensorium Technological Laboratories, Nashville, Tennessee 37210, USA}

\author{\firstname{Otto L.} \surname{Muskens}}
\email[e-mail~: ]{O.Muskens@soton.ac.uk}
\affiliation{Physics and Astronomy, Faculty of Engineering and Physical Sciences, University of Southampton, SO17 1BJ Southampton, UK}

\author{\firstname{Peter R.} \surname{Wiecha}}
\email[e-mail~: ]{pwiecha@laas.fr}
\affiliation{LAAS-CNRS, Universit\'e de Toulouse, UPS, Toulouse, France}

\begin{abstract} 
    Light scattering by spherical-shaped particles of sizes comparable to the wavelength is foundational in many areas of science, from chemistry to atmospheric science, photonics and nanotechnology. With the new capabilities offered by machine learning, there is a great interest in end-to-end differentiable frameworks for scattering calculations. Here we introduce PyMieDiff, a fully differentiable, GPU-compatible implementation of Mie scattering for layered, spherical particles in PyTorch. The library provides native, autograd-compatible spherical Bessel and Hankel functions, vectorized evaluation of Mie coefficients, and APIs for computing efficiencies, angular scattering, and near-fields. All inputs -- geometry, material dispersion, wavelengths, and observation angles and positions -- are represented as tensors, enabling seamless integration with gradient-based optimisation or physics-informed neural networks. The toolkit can also be combined with ``TorchGDM'' for end-to-end differentiable multi-particle scattering simulations. 
    PyMieDiff is available under an open source licence at \url{https://github.com/UoS-Integrated-Nanophotonics-group/MieDiff}.
    \\ \textbf{Keywords:} Mie theory, core‑shell particles, automatic differentiation, gradient optimisation, PyTorch, GPU acceleration
\end{abstract}

\maketitle
\section{Introduction}

Mie theory provides an analytical solution for light scattering by spherical particles, which underpins many nano-optical applications (e.g., color generation, dielectric metamaterials, nanoantennas, radiative cooling).\cite{mieBeitrageZurOptik1908}
Multi-layer (core–shell) spheres extend this to complex nanoparticle designs. While the forward scattering problem can be solved by standard Mie formulae,\cite{bohrenAbsorptionScatteringLight1998} the inverse design of such particles remains challenging: it typically requires many repeated forward simulations and costly optimization.
Recent work has therefore turned to machine learning, training neural nets to predict scattering spectra and act as differentiable surrogates.\cite{kuhnInverseDesignCoreshell2022, peurifoyNanophotonicParticleSimulation2018, soSimultaneousInverseDesign2019,estrada-realInverseDesignFlexible2022, sounGradientbasedOptimizationCoreshell2025}
An alternative approach, which involves integrating the exact Mie solution directly into gradient-based design workflows has not been fully exploited, as analytic derivatives tend to become very bulky,\cite{liComputationMieDerivatives2013} especially for core-shell spheres.

A variety of open-source packages implement Mie theory for spheres.
For example, ``MiePython'' is a pure-Python (NumPy/Numba) implementation of Mie theory for homogeneous spheres.\cite{prahlMiepythonPythonLibrary2025}
``PyMieSim'' is an open-source Python/C++ toolkit that supports scattering from spheres, infinite cylinders and core–shell geometries.\cite{sivry-houlePyMieSimOpensourceLibrary2023}
``Scattnlay'' is a c++ program, focused on the calculation of nearfields inside and around multi-layer spherical particles.\cite{ladutenkoMieCalculationElectromagnetic2017}
``pyMieCS'' is a vectorized NumPy library specialized for core–shell nanoparticles.\cite{wiechaPymiecsSimplePython2024}
These tools offer validated, high-speed forward solvers, but none provide native differentiation capabilities or GPU backends.

Automatic differentiation (AD) is the key numerical technique behind deep learning, and allows calculating arbitrary derivatives for any numerical calculation with close to analytical precision.\cite{wengertSimpleAutomaticDerivative1964}
The photonics community has seen rapid growth of differentiable simulation tools based on AD. It is a generalization of the commonly used adjoint method.\cite{capersDesigningCollectiveNonlocal2021,grieshammer2024continuous,bahmaniTopologyOptimizationOptical2025}
While the adjoint method requires manual, analytical derivation of reciprocal physics to obtain gradients, AD programmatically tracks every operation within a generalized computational graph to enable backward calculation of arbitrary derivatives.
AD provides high-level flexibility at the cost of a significant code complexity and memory overhead, required to store the computational graph. 
Recent efforts have produced open-source Maxwell solvers with automatic differentiation (e.g., FEM, Fourier modal, FDTD or volume integral approaches) which enable gradient-based inverse design and AD-driven topology optimization.\cite{hughesForwardModeDifferentiationMaxwells2019, minkovInverseDesignPhotonic2020, colburnInverseDesignFlexible2021, wangDifferentiableEngineDeep2022, luceMergingAutomaticDifferentiation2024, kimMeentDifferentiableElectromagnetic2024,ponomarevaTorchGDMGPUAcceleratedPython2025, mahlau2026fdtdx}
These tools have been implemented using a range of scripting languages and libraries. The majority of recent tools are written in Python, where currently the most commonly used autodiff libraries are PyTorch and JAX.
For Mie theory computations however, no autodiff capable tool exists so far.

Recent works have highlighted the power of integrating physics-based models with deep learning. An important example are deep learning based nanoparticle design approaches which, so far, often rely on data-based tandem networks or surrogate models.\cite{liuTrainingDeepNeural2018, soSimultaneousInverseDesign2019, estrada-realInverseDesignFlexible2022, khaireh-waliehNewcomersGuideDeep2023}
These solutions often suffer from a lack of surrogate model fidelity \cite{dinsdaleDeepLearningEnabled2021}, as their interpolation is limited to the dataset they are trained on. These large datasets only cover a limited parameter space, and new problems require new datasets.
Replacing such surrogate models by AD capable analytical solvers would suppress this typical point of failure.

We present the PyTorch-based toolkit ``PyMieDiff'' which implements analytical Mie theory for core–shell spherical particles with full support for automatic differentiation.
Our library implements the standard Mie recurrence (spherical Bessel/Hankel forward and backward recurrences, angular functions, vector spherical harmonics\cite{caiComputationSphericalBessel2011}) entirely in PyTorch.
All quantities – particle geometry (layer thicknesses, materials), wavelengths, and scattering angles – can be treated as differentiable tensors.
As a result any Mie-derived observable can be backpropagated to compute gradients with respect to any input parameter.
The code is fully vectorized over Mie orders, wavelengths, angles, and positions. It supports batched evaluation of many particles in parallel to maximize computational efficiency. Thanks to the native pytorch implementation it also runs on GPU.

\begin{figure}[!t]
    \begin{center}
        \includegraphics[width=\columnwidth]{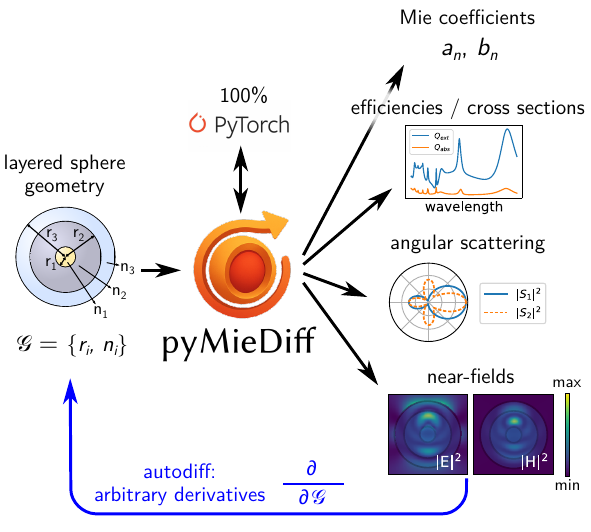}
    \end{center}
    \caption{
        Overview of PyMieDiff's features. 
        The core of the code is a differentiable Mie solver, which is used to compute various differentiable observables (efficiencies, angular scattering, near-fields). All calculations are fully implemented in PyTorch, they all support automatic differentiation and run on GPU.
        The shown results are from a gold-silicon-titania 3-layer particle with core radius $r_1=50\,$nm, inner layer radius $r_2=150\,$nm and outer shell radius $r_2=210\,$nm, placed in vacuum and illuminated by a linearly polarized plane wave.
        The shown spectrum goes from 500\,nm to 1400\,nm. The angular scattering is calculated at 770\,nm and the near-fields at 525\,nm.
        Material dispersion is interpolated from ``refractiveindex.info''.}
    \label{fig:overview}
\end{figure}

Our simple API exposes a \code{Particle} class for defining core-shell spheres with possibly dispersive materials (using the refractiveindex.info format\cite{polyanskiyRefractiveindexinfoDatabaseOptical2024}).
It implements methods to compute extinction, absorption and scattering efficiencies, far-fields and S-matrix elements, as well as near-fields, all with autograd support, as illustrated in figure~\ref{fig:overview}.

Key features of our toolkit include:
\begin{itemize}
    \item Mie coefficients, efficiencies and cross sections, angular scattering, and near-fields for core-shell particles
    \item Compatibility with TorchGDM for multi-particle scattering simulations with autodiff support.\cite{ponomarevaTorchGDMGPUAcceleratedPython2025}
    \item End-to-end auto-differentiability: the entire Mie computation is implemented in PyTorch, allowing gradient-based optimization of particle parameters (sizes, layer thicknesses, refractive indices, etc.) through backpropagation.
    \item GPU-accelerated vectorization: batches of particles, wavelengths and angles are processed in parallel, potentially yielding orders-of-magnitude speedups when workloads are large (especially relative to serial codes).
    \item Flexible materials interface: PyMieDiff supports using refractive index data for real materials (from refractiveindex.info jaml files). Dispersion data is interpolated through a PyTorch-based interpolation routine, so material dispersion enters the gradient graph transparently.
    \item User-friendly API: a high-level object oriented API (\code{Particle} class) as well as a simple to use functional API make it easy to plug the Mie solver into other pipelines, including physics-informed neural networks or design loops.
\end{itemize}

By embedding analytic Mie theory in an autodiff framework, our toolkit enables new design approaches in nano-optics. For example, we demonstrate using the analytic forward model in gradient-based inverse design of core–shell particles (including a ``tandem'' neural network architecture whose forward layer is our Mie solver).
Because the Mie calculations are exact, this avoids the need for approximate surrogate training.\cite{estrada-realInverseDesignFlexible2022} It also opens the door to differentiable sensitivity analyses and rapid parameter retrieval from data.

Potential applications of our autodiff Mie solver include (but are not limited to):

\begin{itemize}
    \item Inverse design and optimization: using gradients to fit particle geometries or material choices to a target scattering spectrum or objective (extinction, backscatter, field enhancement, etc.), even in high-dimensional parameter spaces.
    \item Machine learning integration: embedding the Mie forward model as a layer in neural networks (e.g., tandem or physics-informed networks) so that analytic gradients flow through the scattering physics for training.\cite{jiangMultiobjectiveCategoricalGlobal2020}
    \item Fast batch simulation: efficient GPU-based evaluation of large ensembles of spheres over broad frequency bands for applications like cloud radiative transfer, hyperspectral imaging design, or Monte Carlo light propagation.
    \item Adjoint sensitivity analysis: computing how scattering observables change with small perturbations in geometry or material (for uncertainty quantification or experimental fitting), leveraging the computed autograd derivatives.
\end{itemize}

In summary, our PyTorch Mie toolkit provides a fully autodifferentiable, vectorized and GPU-accelerated implementation of core–shell Mie theory.
This bridges analytic nanophotonics and modern ML tools, enabling gradient-based design methods that were previously difficult or impossible with black-box Mie solvers.
While we demonstrate the toolkit on optical-frequency examples, the implementation is completely general and can be applied at any wavelength (from microwaves to X-rays) where Mie scattering is relevant.

\section{Implementation}

Mie theory describes the interaction of light with a spherical particle through field expansion with spherical waves, using the coefficients $a_n, b_n, c_n$ and $d_n$, where $a_n$ and $b_n$ are the coefficients for the outgoing (scattered) fields, while $c_n$ and $d_n$ are the expansion coefficients for the incoming fields (required in the expansion of internal fields).

In the case of a layered particle, the Mie coefficients are functions of the scale factors $x_l=k r_l$ (with wavevector $k=k_0 n_{\text{env}}$; $k_0 = 2\pi / \lambda_0$; and radius $r_l$ of the $l$th spherical layer) and the relative, complex refractive indices $m_l=n_l / n_{\text{env}}$ for layer $l$.
The explicit form of the Mie scattering coefficients is given in the appendix~\ref{app:mie_coeff}.
The evaluation of the Mie coefficients requires Riccati-Bessel functions and their derivatives:
\begin{equation}
    \psi_{n}(z)=z j_{n}(z), \quad \xi_{n}(z)=zh_{n}^{(1)}(z), \quad \chi_n(z)=-zy_n(z).
\end{equation}
where $j_n$ and  $y_n$ are spherical Bessel functions of the first, respectively second kind, and $h^{(1)}_n=j_n + \text{i} \, y_n$ are spherical Hankel functions of the first kind.

Implementing Mie series directly based on these spherical Bessel functions can become numerically unstable for large size parameters or absorbing materials.\cite{wiscombeImprovedMieScattering1980} 
Therefore, modern implementations reformulate the Mie expansion using logarithmic derivatives of the form 
\begin{equation}\label{eq:logderiv_form}
    D_n^{(1)}(z) \;=\; \frac{\psi_n' (z)}{\psi_n (z)}\, .
\end{equation}
These can be numerically calculated through similar recurrences as for the normal spherical Bessel functions. An important detail is that upward recurrences are stable for the logarithmic derivatives of $\xi_n$, while downward recurrences are needed to calculate the logarithmic derivatives for $\psi_n$.\cite{yangImprovedRecursiveAlgorithm2003}
All mathematical and algorithmic details can be found in the appendix~\ref{app:mie_coeff}. Our implementation closely follows the work of Pe\~na and Pal.\cite{penaScatteringElectromagneticRadiation2009}

In PyTorch (as of version 2.10), neither spherical Bessel functions, nor their logarithmic derivatives are implemented.
The key contribution of our work is therefore to provide PyTorch-based, AD-capable, vectorized, fast and stable spherical Bessel routines.
We implement following AD enabled functions:
(1) A PyTorch wrapper to the SciPy versions of standard spherical Bessel functions and their derivatives.
(2) A native PyTorch implementation of spherical Bessel functions and their derivatives based on up- and down-recurrences.
(3) A native PyTorch implementation of logarithmic derivatives and all further ingredients to implement the stable Mie algorithm, suggested by Yang and further improved by Pe\~na and Pal.\cite{yangImprovedRecursiveAlgorithm2003, penaScatteringElectromagneticRadiation2009}
The spherical Bessel and Hankel functions were mainly implemented for testing. But since they may be useful for various other applications that require torch-versions of these functions with native GPU support, all special functions are available through the ``pymiediff.special'' module.

Based on the logarithmic derivatives, PyMieDiff implements a native PyTorch Mie theory stack comprising Mie coefficients, angular functions, and vector spherical harmonics. This is then used to implement functions that evaluate crossections, angular radiation, and near-fields.

As ``PyMieDiff'' is entirely implemented in PyTorch, automatic differentiation is fully supported through all calculations.

\begin{figure}[!t]
    \begin{center}
        \includegraphics[width=\columnwidth]{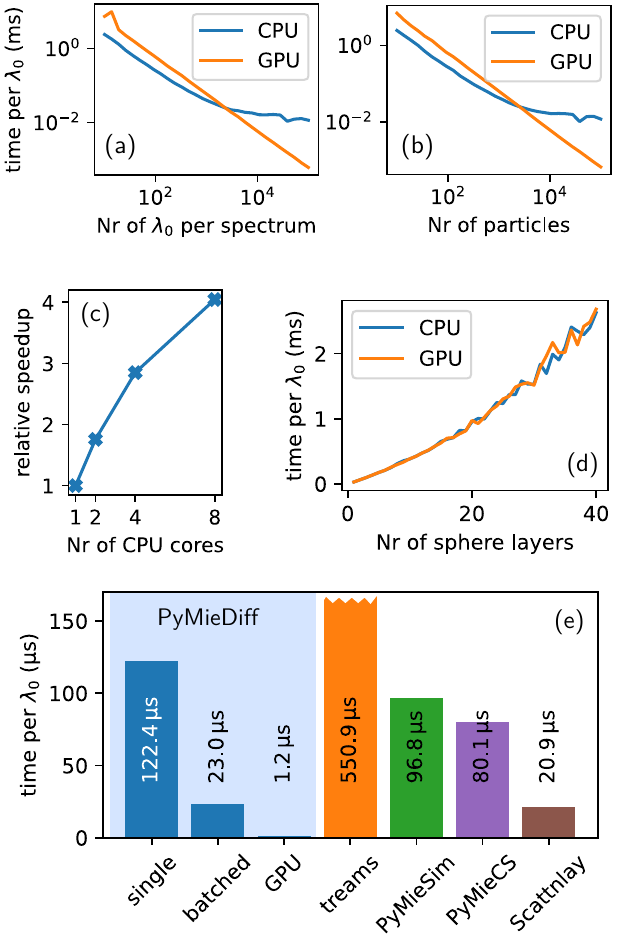}
    \end{center}
    \caption{
        Benchmarks. If not otherwise noted, benchmarks were run for a single dielectric core-shell particle at 256 wavelengths.
        (a) timing for increasing numbers of simultaneously evaluated wavelengths.
        (b) timing for increasing numbers of simultaneously evaluated particles (single wavelength).
        (c) parallelization speedup on multi-core CPUs.
        (d) timing for increasing nr. of shell layers.
        (e) timing comparison to other Mie toolkits (on CPU). The cases ``batched'' and ``GPU'' simultaneously evaluate 256 different particles, at 256 wavelengths each.
        Our CPU is a Zen 3 AMD processor. Our GPU is an NVIDIA RTX 4090.
        Note that ``treams'' is not performance-optimized for Mie evaluation since it is not a dedicated Mie toolkit (but T-matrix), in consequence the time-bar for the treams result is not fully shown.
    }
    \label{fig:benchmark}
\end{figure}

\section{Benchmark}

In figure~\ref{fig:benchmark} we assess the computational performance of PyMieDiff and compare it against existing Mie scattering toolkits. 
Owing to its fully vectorized implementation utilizing PyTorch tensors, PyMieDiff exhibits efficient scaling for any independent parameter, like number of wavelengths per spectrum (Fig.~\ref{fig:benchmark}a), scattering angles, evaluation positions, the number of particles (Fig.~\ref{fig:benchmark}b).
The number of layers requires evaluation of a cascade of interdependent scattering coefficients (see appendix~\ref{app:mie_coeff}), thus it scales linearly with compute (Fig.~\ref{fig:benchmark}d).
We observe that the evaluation time is dictated by memory transfer until around the order of $10^3$ parallel evaluations.
Running PyMieDiff on GPU shows that the performance is memory-bound until at least $10^5$ concurrent Mie evaluations. GPU becomes advantageous only for large batch sizes above $\approx 10^3$ (see figures~\ref{fig:benchmark}a-b).
But even for lower batch sizes, GPU support can still benefit integration into deep learning schemes, as memory transfer during model evaluation is reduced and the neural network parts of the calculation may strongly benefit from GPU evaluation also on smaller batch sizes.

\begin{figure}[!t]
    \begin{center}
        \includegraphics[width=0.95\columnwidth]{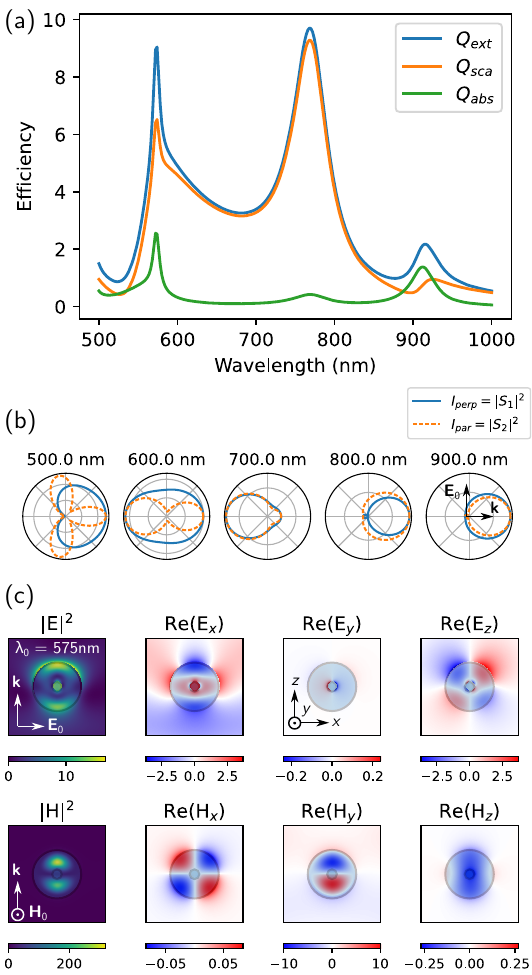}
    \end{center}
    \caption{Mie forward evaluation by the example of scattering from a gold-silicon core-shell particle with core radius $r_c=20\,$nm and shell radius $r_s=100\,$nm, placed in vacuum and illuminated by a linearly polarized plane wave. 
    Tabulated material permittivities are taken from literature.\cite{johnsonOpticalConstantsNoble1972, edwardsSiliconSi1997}
    (a) Extinction (blue), scattering (orange), and absorption (green) efficiency spectra.
    (b) Scattering radiation patterns in the scattering plane at selected wavelengths for perpendicularly (blue lines) and parallel (orange dashed lines) polarized light.
    (c) Near-fields (electric and magnetic in top and bottom row) inside and around the particle, evaluated at the resonance at $\lambda_0=575\,$nm. 
    The shown maps are $400\times 400\,$nm$^2$. Colorbars indicate field intensity (leftmost panels) and amplitude (second from left to right panels), relative to the incident field absolute amplitude. 
    }
    \label{fig:example_forward}
\end{figure}

\begin{lstlisting}[language=Python, float, caption={Particle class usage example for forward Mie calculations.}, label={lst:particle_class}]
import torch
import pymiediff as pmd

# --- setup
r_core = 20.0  # nm
r_shell = 100.0  # nm
mat_core = pmd.materials.MatDatabase("Au")
mat_shell = pmd.materials.MatDatabase("Si")
n_env = 1.0

p = pmd.Particle(
    r_layers=[r_core, r_shell],
    mat_layers=[mat_core, mat_shell],
)

# --- evaluation
wl0 = torch.linspace(500, 1000, 50)  # nm
k0 = 2 * torch.pi / wl0

# efficiency spectra
cs = p.get_cross_sections(k0)

# angular scattering (angles in radian)
theta = torch.linspace(0.0, 2 * torch.pi, 100)
angular = p.get_angular_scattering(k0, theta)

# nearfields (positions in nm)
x, z = torch.meshgrid(
    torch.linspace(-250, 250, 50),
    torch.linspace(-250, 250, 50),
)
y = torch.ones_like(x)
r_probe = torch.stack([x, y, z], dim=-1)

fields = p.get_nearfields(k0=k0, r_probe=r_probe.view(-1, 3))

\end{lstlisting}

On multi-core CPUs, PyMieDiff demonstrates good parallel efficiency, slowing down for eight or more cores (Fig.~\ref{fig:benchmark}d). We attribute this reduction in scaling efficiency to memory transfer overhead in PyTorch's shared-memory parallelization, not optimized for embarrassingly parallel tasks.
We note that these parallelization capabilities are automatically used since they are inherent to PyTorch. 
This scaling behaviour is particularly beneficial for large parameter sweeps, many-particle optimization tasks or machine learning integration with batched training schemes.
In a direct comparison with other publicly available toolkits (Fig.~\ref{fig:benchmark}d), PyMieDiff is on par with the fastest implementation (scattnlay) and achieves markedly lower evaluation times per wavelength as soon as batched GPU evaluation is possible. 
At best, our toolkit provides more than one order-of-magnitude improvement over the fastest of the other evaluated tools. 
Note that while the T-Matrix package ``treams''\cite{beutelTreamsTmatrixbasedScattering2024} is included for completeness, it is not parallelized and not specifically optimized for single particle Mie calculations.

Overall, these benchmarks establish PyMieDiff as a performant and scalable solution for Mie scattering computations, combining differentiability, GPU compatibility, and multi-core efficiency in a single framework.

\begin{figure}[!t]
    \begin{center}
        \includegraphics[width=\columnwidth]{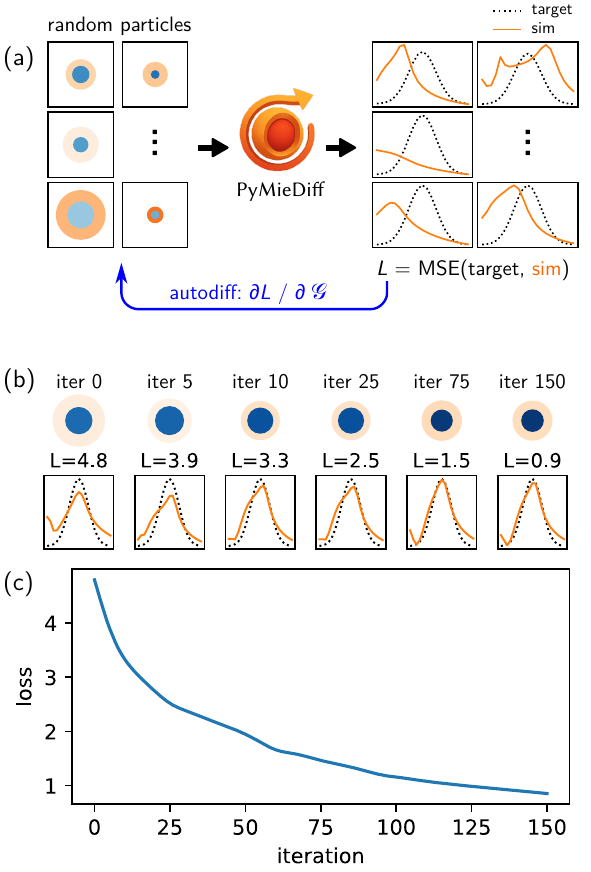}
    \end{center}
    \caption{
        Example of gradient-based particle optimization.
        (a) Sketch of the particle optimization procedure. A large number of random core-shell particles is randomly initialized and evaluated using PyMieDiff. Their optical properties (orange lines, here: scattering efficiency spectra) 21 points between $400\,$nm and $800\,$nm are compared with the design target (dotted black lines, here: a Gaussian response). The derivatives of the loss function (here MSE) with respect to the geometry parameters are obtained via PyTorch autograd and used to iteratively update the geometries.
        (b) Sketches of the best particles of selected iteration during an optimization (top row) together with their spectral response (bottom row). Relative sizes are on the same scale, the shades of the colors indicate the real part of the refractive index (darker means higher).
        (c) Loss function during the optimization.
    }
    \label{fig:example_optimization}
\end{figure}

\begin{lstlisting}[language=Python, float, caption={Automatic differentiation usage example.}, label={lst:autodiff}]
# - gradient of scattering wrt wavelength
wl = torch.as_tensor(500.0)  # nm
wl.requires_grad = True
cs = p.get_cross_sections(k0=2 * torch.pi / wl)

cs["q_sca"].backward()
dQdWl = wl.grad
\end{lstlisting}

\section{Examples}

\subsection{Forward evaluation}\label{sec:forward}

As a first example, we demonstrate the forward solver by simulating the extinction, scattering and absorption efficiencies, as well as angular scattering patterns of a gold-silicon core-shell particle.
The efficiency spectra are shown in figure~\ref{fig:example_forward}a, scattering patterns of perpendicular and parallel polarized light are given in Fig.~\ref{fig:example_forward}b.
Electric and magnetic near field intensities and real parts of all field components are shown in Fig.~\ref{fig:example_forward}c.

The code to configure the particle and calculate the spectra, angular scattering patterns and near-fields is given in listing~\ref{lst:particle_class}, where tabulated materials are handled by the \code{MatDatabase} class and the \code{Particle} class acts as an easy-to-use high-level interface to PyMieDiff.

We compared the results to several other, openly available Mie solvers and obtained results identical to machine precision. 
This comparison can be found in the examples of the online documentation.

\begin{lstlisting}[language=Python, float, caption={Example how to use autodiff for a simple optimization.}, label={lst:gradient_optimization}]
import torch
import pymiediff as pmd

# setup
wl0 = torch.tensor([700.0])  # nm
k0 = 2 * torch.pi / wl0
mat_particle = pmd.materials.MatDatabase("Si")
n_env = 1.0

# optimization
r_init = 60.0  # initial guess
r_opt = torch.tensor([r_init], requires_grad=True)
optimizer = torch.optim.Adam(params=[r_opt], lr=0.5)

# gradient loop
for i in range(100):
    optimizer.zero_grad()
    particle = pmd.Particle(
        mat_env=n_env,
        r_layers=r_opt,
        mat_layers=[mat_particle],
    )
    q_sca = particle.get_cross_sections(k0)["q_sca"]

    loss = -q_sca  # *maximize* scattering: minus sign
    loss.backward()  # calc. gradients via autodiff
    optimizer.step()  # apply gradients on params

print("radius for strongest scattering:", r_opt.item())

\end{lstlisting}

\begin{figure*}[!t]
    \begin{center}
    \includegraphics[width=0.95\linewidth]{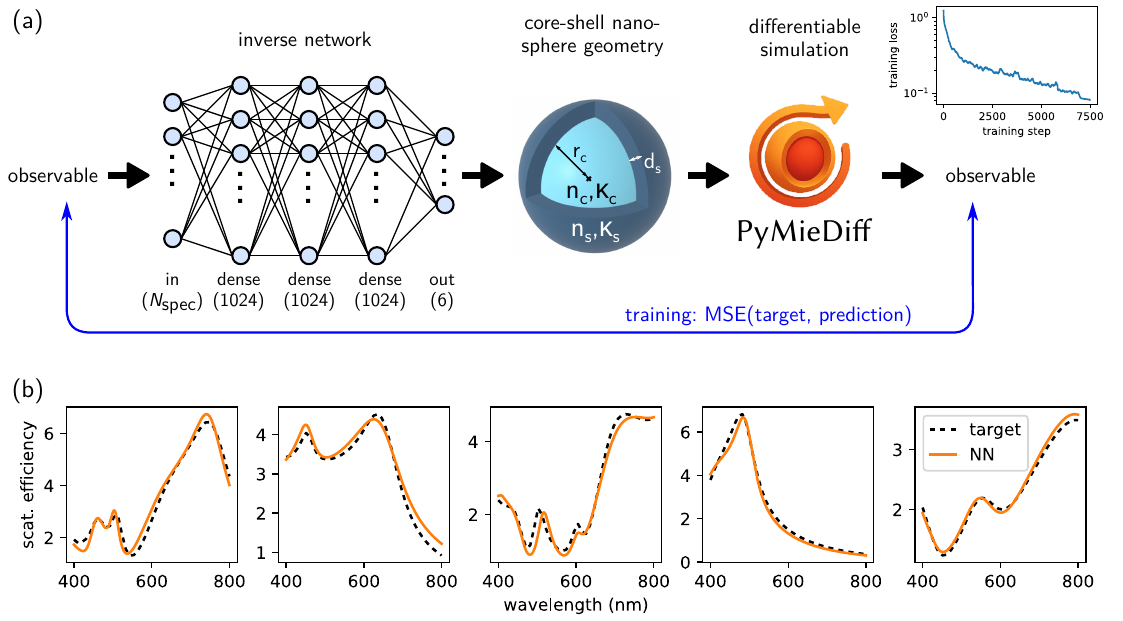}
    \end{center}
    \caption{
        ``Mie-informed'' tandem network.
        (a) Sketch of the training configuration. The design observable (e.g., scattering efficiency, or any other Mie-calculated value) is fed into the ``inverse network'' (here a simple MLP). The NN result is interpreted as core-shell geometry parameters and fed into PyMieDiff, which calculates the response of the particle analytically. The result of the Mie calculation is compared to the original input data using some loss function (here MSE). The inverse network is trained using backpropagation of the loss through the differentiable Mie solver and then the network itself.
        The inset shows the training loss for an inverse network trained on scattering efficiency spectra of dielectric particles.
        (b) Examples of inverse design responses (orange lines) vs. randomly sampled design targets (dashed black lines) from the training dataset.
        Spectral matching is done at $40$ points over the wavelength range between $400$\,nm  and $800$\,nm.
        }
\label{fig:example_tandem}
\end{figure*}

\subsection{Gradient based optimization}\label{sec:optimization}

Design problems in nanophotonics are often non-trivial inverse problems that require numerical optimization. Here, we will try to find a core-shell particle with a response as close as possible to a hand-drawn spectrum. Since no particle likely has exactly these properties, the problem is ill-posed. 
Furthermore, many high-dimensional problems are naturally ill-posed, in these cases rather because several possible solutions exist.\cite{khaireh-waliehNewcomersGuideDeep2023}
Global optimization is useful in cases with many local minima or for non-continuous / discrete problems.
Using local gradient based optimization on a large number of initial guesses provides a straightforward alternative with very fast convergence.\cite{dengNeuraladjointMethodInverse2021}

The automatic differentiation capability of PyMieDiff is therefore the key contribution of this work.
In the following example we demonstrate how to use autodiff for optimization of a core-shell particle to implement as closely as possible a predefined Gaussian scattering response.

We start by defining a loss function, using the mean square error (MSE) of the current iteration's particle responses vs. the target response:
\begin{equation*}
        L = \frac{1}{n} \sum_{i=1}^{n}\left(R_{\text{target},i} - R_{i}\right)^{2}\, .
\end{equation*}
Here the $R_i$ serve as general placeholders for observables calculated by Mie theory, and $n$ is the total number of calculations (e.g., multiple wavelengths, angles, particles). The optimization targets are given by $R_{\text{target},i}$. 
Via automatic differentiation, we calculate the partial derivatives of $L$ with respect to the geometric parameters $\partial L / \partial \mathcal{G}$, and update the particles to minimize $L$. To avoid getting stuck in local minima, we can perform this optimizaiton on a large number of initial guesses. 
This optimization procedure is depicted in figure~\ref{fig:example_optimization}a.

In this example, we optimize the scattering efficiency $Q_{\text{sca}}$ of a dielectric core-shell particle over $400-800$\,nm to match the Gaussian centred at $600$\,nm. Its core and shell size as well as the material refractive indices are varied. We limit the radii to $10-100$\,nm and the complex refractive indexes to dielectric materials with $n_{real}$ ranging from 1 to 4.5 and $n_{imag}$ from 0.0 to 0.1. 
To keep the optimization within these restrictions, we optimize normalized values using a sigmoid activation inside the optimization loop.

In this example, the popular Adam optimizer is used.\cite{kingmaAdamMethodStochastic2014} 
While Adam is ideal for mini-batch based optimization as typically used for neural network training, it performs well also on our optimization task. 
We demonstrate in the online documentation how to use the L-BFGS optimizer as an alternative.\cite{nocedalUpdatingQuasiNewtonMatrices1980}

We perform a batch optimization on 100 random core-shell particles in parallel, which takes roughly $100-200$\,ms per iteration on a typical office CPU.
The spectra from the iteratively improved solutions are shown in figure~\ref{fig:example_optimization}b alongside the target spectra.
The convergence curve is shown in figure~\ref{fig:example_optimization}c. 
We also found that higher learning rates are usually stable, allowing convergence within some 20-30 iterations. Yet, as optimization is fast, lower, very stable learning rates can be used without major inconvenience.

Listing~\ref{lst:gradient_optimization} gives a simple code example, demonstrating how a gradient based optimization can  be implemented with PyMieDiff.

\subsection{Core-shell design: Mie informed tandem model}\label{sec:tandem}

Our toolkit being entirely implemented in Pytorch, it can be directly implemented within any existing PyTorch machine learning pipeline. Here we demonstrate how to train a deep learning neural network through a Mie-theory based loss function.

As a simple technical example, we train a design network capable of predicting core-shell geometries that fulfil given target optical properties. This so called ``tandem'' model is a commonly used deep learning method for inverse design.\cite{liuTrainingDeepNeural2018, khaireh-waliehNewcomersGuideDeep2023}
It works by regularizing the training of the ill-posed inverse problem through a forward predictor, as depicted in figure~\ref{fig:example_tandem}a. 
Typically, the forward model is a trained neural network that approximates a physics solver, adding automatic differentiation capabilities.
With PyMieDiff as a fast, autograd-compatible Mie solver, analytical Mie theory can be used as an error-free forward model through which the loss can be backpropagated to train the inverse network.

\begin{figure*}[!t]
    \begin{center}
    \includegraphics[width=0.95\linewidth]{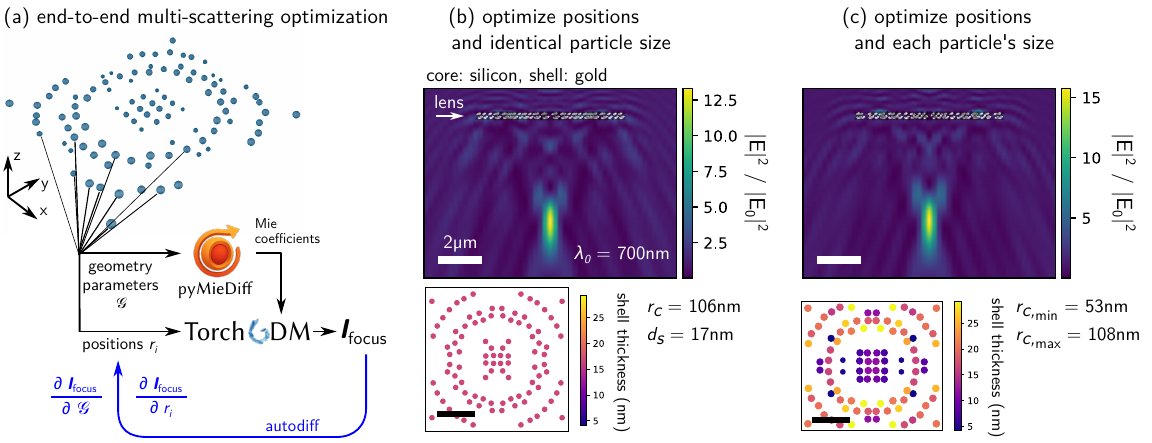}
    \end{center}
    \caption{
        Combining PyMieDiff with torchGDM\cite{ponomarevaTorchGDMGPUAcceleratedPython2025} for end-to-end multi-scattering autograd calculations.
        (a) Sketch of the gradient calculation scheme. PyMieDiff is used to calculate differentiable Mie coefficients, which are in turn used to create TorchGDM structure models for multi-particle scattering simulations. Gradients can then be obtained with respect to all input parameters, including particle geometry and material properties (represented by $\mathcal{G}$) and particle positions ($r_i$).
        (b) Optimization of a diffractive lens, composed of identical particles. Positions as well as the size parameters of the particle are optimized. Top: field intensity enhancement in side view, bottom: top view of the geometry.
        (c) Same as (b), but now each particle's core and shell sizes are optimized individually.
        Core sizes are indicated by the circle size, shell size by the color code. Wavelength is $\lambda_0=700$\,nm. All scale bars are 2\,$\mu$m. 
    }
\label{fig:example_torchgdm}
\end{figure*}

We train the Mie-informed tandem network on the same type of particles as in the optimization example, limiting the materials to constant dielectrics and training the model on spectra in the visible light range ($400-800$\,nm). 
A dataset of 20,000 spectra, each consisting of 41 wavelength points, calculated from randomized particles, is used for training the model.
The inverse model consists of a fully connected neural network (multilayer perceptron, MLP) with three hidden layers, each containing 1024 neurons, connected by ReLU activation functions. 
A sigmoid function is used at the output layer to ensure positive predictions within our defined particle limits. 
Using a basic manually optimized training schedule, with two learning rates (first $10^{-4}$, then $10^{-5}$) and incremental increase of the batch size (from 32 up to 256 in four steps), we obtain fast convergence after a few thousand training steps (see inset in Fig.~\ref{fig:example_tandem}a). The training takes a few minutes on an 8-core office CPU.

Once the inverse model is trained, we use scattering spectra from particles that were not in the training set as test targets and predict the geometry parameters by the inverse network. 
A few examples are shown in figure~\ref{fig:example_tandem}b, demonstrating the model's performance.
By comparing predicted core-shell particles to the reference particles used as design target, the results of the tandem network could now for example be used to assess ``how ill posed'' the studied problem of core-shell design is. The more non-unique solutions exist, the larger the discrepancy between target particles and predicted designs.
Note that, as the forward model is ``perfect'' (exact Mie solution), the discrepancy of the solutions with respect to the target spectra comes solely from the inverse network, for which we deliberately use a very simple dense network, chosen to have a short training time of only a few minutes.

\subsection{Autodiff multi-scattering: End-to-end metasurface design}\label{sec:combine_with_torchgdm}

As a final example, we demonstrate how PyMieDiff can be combined with the autodiff light scattering simulation toolkit ``TorchGDM'',\cite{ponomarevaTorchGDMGPUAcceleratedPython2025} to perform multi-particle scattering simulations. 
This combination enables similar functionality as offered by the T-matrix method, implemented for instance in frameworks such as CELES,\cite{CELES2017103} SMARTIES,\cite{smarties201639} or treams.\cite{beutelTreamsTmatrixbasedScattering2024}
However, our approach enables full automatic differentiation through multi-scattering simulations.

PyMieDiff evaluates the Mie response of one or several spherical particles, which are then used by TorchGDM to generate a structure model and calculate the optical response of an ensemble of many scatterers, where all optical interactions between the particles are taken into account. 
Since both tools fully support PyTorch automatic differentiation, gradient optimization can be used to iteratively optimize the positions and core/shell size parameters for any target observable. 
This is illustrated in figure~\ref{fig:example_torchgdm}a.

We demonstrate this capability on a simple toy problem, not meant to be relevant for an actual application. 
We calculate the field intensity enhancement at a target focal position, to optimize a diffractive lens made of silicon-gold core-shell particles and optimize for the core and shell radii and the particle positions on the XY plane.
Starting from the same initial conditions ($10\times 10$ identical 90\,nm/15\,nm silicon/gold core/shell spheres on a regular grid), we compare two scenarios: (1) Optimization of the positions of many identical particles, illustrated in figure~\ref{fig:example_torchgdm}b, and (2) optimization of particle positions and each particle's core and shell size, illustrated in figure~\ref{fig:example_torchgdm}c.
As expected, we obtain a higher focal intensity if more degrees of freedom are available (individual particle sizes).

The multi-scattering capability may be useful, for instance, in the design of bottom-up metasurfaces or for fitting optical scattering data from dense, spherical solutions, where homogenization models fail.

\section{Limitations and perspectives}\label{sec:limitations}
Here we list current limitations of PyMieDiff and possible future extensions:
\begin{itemize}
\item PyMieDiff is currently limited to isotropic materials and spherical (3D) particles. 
In the future we may implement support for tensorial permittivities as well as 2D Mie theory for multi-shell infinite cylinders.
\item Vector spherical harmonics are currently implemented only for order $l=1$, which limits the field calculations to particles with spherical symmetry. 
We plan to add pure PyTorch implementations of general vector spherical harmonics that could be used for field calculations of general T-matrices. 
A possible route could be to expand on NVIDIA's recent work about spherical Fourier neural operators.\cite{bonevSphericalFourierNeural2023}
\item While the logarithmic derivatives algorithm is very stable, there is one limitation in combination with vectorization: 
In a single batch of many concurrent evaluations, the Mie order required for the largest size parameter case is used for all calculations in that batch.
This can in fact become unstable for \emph{small} particles, causing small size parameter cases in the batch to overflow and yield NaN results (``Not a Number'') for these cases. 
While this is not ideal, it is failing with a non-numerical result and will directly be noted by the user. 
A simple split of the batch into a smaller and a larger fraction suffices to make it work.
\end{itemize}

\section{Conclusions}
In this work, we present PyMieDiff, an implementation of Mie scattering for layered spherical particles, fully built in the automatic differentiation framework PyTorch. 
It is available as open source software at \url{https://github.com/UoS-Integrated-Nanophotonics-group/MieDiff}.
The package enables gradient-based optimisation and hybrid physics-informed deep learning models. 
It can be combined with other PyTorch-based toolkits, such as torchGDM, to perform end-to-end differentiable multi-particle scattering simulations.
The toolkit was designed with flexibility and performance in mind, offering a native PyTorch implementation of the entire Mie algorithm, with full GPU support.

We demonstrate the capabilities by several examples of inverse design problems, including the iterative reconstruction of core-shell particle geometries from a target scattering spectrum, neural network training through analytical Mie calculations, and the gradient-based design of a diffractive lens made of several core-shell spheres, for which PyMieDiff is combined with the multi-particle scattering toolkit ``TorchGDM''.

At the time of the submission, we became aware of a manuscript developing similar ideas.\cite{asadovaGradientbasedOptimizationScatterer2025} This underpins the timeliness and importance of the differentiable formulation of algorithms that solve multiple-scattering problems, a key requirement for inverse design of photonic nanostructures.

\begin{acknowledgments}
    OJ acknowledges support by EPSRC through a PhD studentship. 
    OLM acknowledges support by EPSRC through Grant No. EP/W024683/1.
    PRW acknowledges financial support by the French Agence Nationale de la Recherche (ANR) under grants ANR-22-CE24-0002 (project NAINOS) and ANR-23-CE09-0011 (project AIM).
    SDL acknowledges financial support under the National Recovery and Resilience Plan (NRRP), Mission 4, Component 2, Investment 1.1, Call for tender No. 1409 published on 14/09/2022 by the Italian Ministry of University and Research (MUR), funded by the European Union – NextGenerationEU – Project Title MINAS - CUP B53D23028420001 - Grant Assignment Decree No. 1380 adopted on 01/09/2023 by the Italian Ministry of University and Research (MUR).
\end{acknowledgments}

\section*{Data Availability}
The supporting data used in this work are openly available from the University of Southampton repository at \url{https://doi.org/10.5258/SOTON/D3776}

\section*{Appendix}

\setcounter{figure}{0}
\makeatletter
\renewcommand{\thefigure}{A\arabic{figure}}
\def\thesection       {}
\def\p@section        {}
\makeatother

\subsection{Stability on very large spheres}\label{app:large_spheres}

\begin{figure}[!t]
    \begin{center}
        \includegraphics[width=\columnwidth]{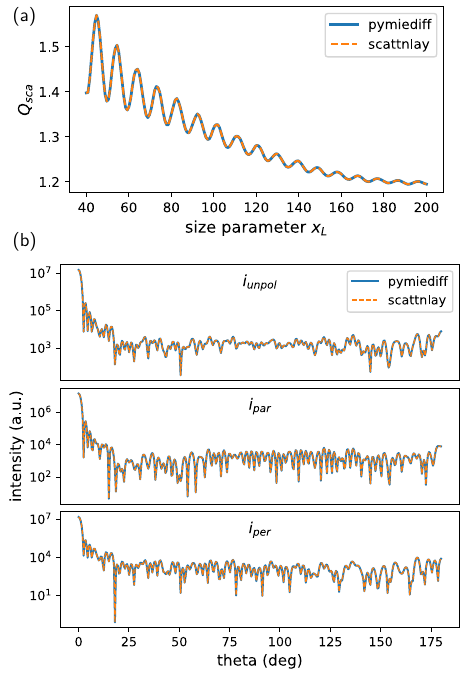}
    \end{center}
    \caption{
        Demonstration of stability on large spheres compared to scattnlay.\cite{penaScatteringElectromagneticRadiation2009}
        (a) Scattering of a soot-coated water sphere in vacuum for increasing size parameter. 
        Used refractive indixes are $n_{\text{water}}=1.33$ and $n_{\text{soot}}=1.59+0.66\,\text{i}$. The soot volume fraction is $0.01$.
        (b) Angular scattering from a layered sphere with layer radii $[135, 2365, 2395, 15000]$\,nm and layer refractive indices $[2.1 + 0.15\text{i}, 1.75, 0.45 + 5.06\text{i}, 3.62]$ at $\lambda_0=1100$\,nm.
        }
    \label{fig:app_lage_sphere}
\end{figure}

To assess the stability of PyMieDiff on very large problems, we reproduce the test performed by Pe\~na and Pal,\cite{penaScatteringElectromagneticRadiation2009} who simulate a soot shell around a large water sphere for increasing size parameters.
The result is shown in Figure~\ref{fig:app_lage_sphere}a.

In Figure~\ref{fig:app_lage_sphere}b we show angular scattering for a 4-layer particle of $30\,$\textmu m outer diameter, consisting mainly of dielectric material but containing also absorbing shells. The size parameter of this problem is 310.

\subsection{Example code}\label{app:example_code}
All examples shown in this work, further examples, as well as the full technical documentation of the package are available online at \url{https://uos-integrated-nanophotonics-group.github.io/MieDiff/index.html}.

\subsection{Vectorization conventions}\label{app:vectorization}

PyMieDiff routines are implemented along the following conventions:

\paragraph*{Mie orders:} The (Mie) order $n$ is passed as an integer specifying the \textit{maximum} order. 
Then, all PyMieDiff routines (special functions, Mie coefficient routines, etc...) calculate all orders up to the maximum order. An additional, order dimension is added as the first dimension to the shape of the passed argument(s).

\paragraph*{Vectorization dimensions: Internal vectorization conventions are the following:} 
\begin{itemize}
    \item First dim.: Order / Mie order $n$ (an additional, first dimension is added upon function calls)
    \item Second dim.: Layer number $l$: $l=1,...,L$
    \item Third dim.: Number of particles $N_{\text{particles}}$
    \item Fourth dim.: Number of wavenumbers $k_0$: $N_{\text{k0}}$
    \item Fith dim.: Number of angles or number of positions (angular scattering / fields)
\end{itemize}

Further dimensions (e.g., for field components) can be added after these four main dimensions and will be broadcast automatically.

\subsection{Riccati–Bessel recurrences}\label{app:RiccatiBessel_recurrences}

To ensure numerical stability, we implement the logarithmic derivatives of the Riccati–Bessel functions, $D_{n}^{(1)}(z)$ and $D_{n}^{(3)}(z)$, rather than their standard derivative, $\psi_n^{\prime}(z)$ and $\xi_n^{\prime}(z)$ \cite{yangImprovedRecursiveAlgorithm2003}. The logarithmic derivative of the first kind, $D_{n}^{(1)}(z)=\psi_n^{\prime}(z) / \psi_n(z)$, is evaluated using the downward recurrence \cite{wiscombeImprovedMieScattering1980}:
\begin{align}
D_{n-1}^{(1)}(z) = \frac{n}{z} - \frac{1}{D_n^{(1)}+\frac{n}{z}}, \quad n=N_{max},\dots,1,
\end{align}
with the initial condition:
\begin{align}
D_{N_{max}}^{(1)}(z) = 0 + i0.
\end{align}
Conversely, the logarithmic derivative of the third kind, $D_n^{(3)}(z)=\xi_{n}^{\prime}(z) / \xi_{n}(z)$, stability across all values of z is achieved by utilizing an upward recurrence relation \cite{Mackowski19901551}:
\begin{align}
D_n^{(3)}(z) = D_n^{(1)}(z) + \frac{i}{\psi_n(z)\xi(x)}, \quad n=1,\dots,N_{max},
\end{align}
This requires an auxiliary upward recurrence to compute the product $\psi_{n}(z) \xi_{n}(z)$:
\begin{align}
\psi_{n}(z) \xi_{n}(z)=\psi_{n-1}(z) \xi_{n-1}(z)\left[\frac{n}{z}-D_{n-1}^{(1)}(z)\right]\left[\frac{n}{z}-D_{n-1}^{(3)}(z)\right],
\end{align}
which is initialized as follows:
\begin{align}
D_0^{(3)}(z)=i,\\
\psi_{0}(z) \xi_{0}(z)=\frac{1}{2}[1-(\cos 2 a+i \sin 2 a) \exp (-2 b)].
\end{align}
The plain Riccati–Bessel functions are found using the recurrence relations \cite{Mackowski19901551,wiscombeImprovedMieScattering1980}:

\begin{align}
\psi_{n}\left(z\right)=\psi_{n-1}\left(z\right)\left[\frac{n}{z}-D_{n-1}^{(1)}\left(z\right)\right], \quad n=1, \ldots, N_{\max },
\end{align}
\begin{align}
\xi_{n}\left(x_{L}\right)=\xi_{n-1}\left(z\right)\left[\frac{n}{z}-D_{n-1}^{(3)}\left(z\right)\right], \quad n=1, \ldots, N_{\max },
\end{align}
starting from,
\begin{align}
\psi_{0}\left(z\right)=\sin \left(z\right), && \xi_{0}\left(z\right)=\sin \left(z\right)-i \cos \left(z\right),
\end{align}
respectfully.
The ideal $N_{max}$ and $N_{stop}$ are found programmatically based of the work of Wiscombe.\cite{wiscombeImprovedMieScattering1980}

\subsection{Vector spherical harmonics}\label{app:vsh}

The vector spherical harmonics $\mathbf{N}_{lm}^{(k)}$ and $\mathbf{M}_{lm}^{(k)}$ can be used to describe the near fields of Mie scattering.\cite{asadovaTmatrixRepresentationOptical2025}
For spherical symmetry, only the $l=1$ odd and even order harmonics are required, which can be written using the Riccati–Bessel functions and their logarithmic derivatives:\cite{ladutenkoMieCalculationElectromagnetic2017}

\begin{align}
\mathbf{N}_{o1n}^{(j)} = \begin{bmatrix}
\sin \phi \; n(n+1) \sin \theta \; \pi_{n}(\cos \theta) \frac{z_{n}(\rho)}{\rho^{2}} \\
\sin \phi \; \tau_{n}(\cos \theta) \frac{D_{n}^{(j)}(\rho) z_{n}(\rho)}{\rho} \\
\cos \phi \; \pi_{n}(\cos \theta) \frac{D_{n}^{(j)}(\rho) z_{n}(\rho)}{\rho}
\end{bmatrix},
\\
\mathbf{N}_{e1n}^{(j)} = \begin{bmatrix}
\cos \phi \; n(n+1) \sin \theta \; \pi_{n}(\cos \theta) \frac{z_{n}(\rho)}{\rho^{2}} \\
\cos \phi \; \tau_{n}(\cos \theta) \frac{D_{n}^{(j)}(\rho) z_{n}(\rho)}{\rho} \\
-\sin \phi \; \pi_{n}(\cos \theta) \frac{D_{n}^{(j)}(\rho) z_{n}(\rho)}{\rho}
\end{bmatrix},
\end{align}
\begin{align}
\mathbf{M}_{o1n}^{(k)} = \begin{bmatrix}
0 \\ 
\cos \phi \; \pi_{n}(\cos \theta) \frac{z_{n}(\rho)}{\rho} \\
-\sin \phi \; \tau_{n}(\cos \theta) \frac{z_{n}(\rho)}{\rho}
\end{bmatrix},
\\
\mathbf{M}_{e1n}^{(k)} = \begin{bmatrix}
0 \\ 
-\sin \phi \; \pi_{n}(\cos \theta) \frac{z_{n}(\rho)}{\rho}  \\
-\cos \phi \; \tau_{n}(\cos \theta) \frac{z_{n}(\rho)}{\rho}
\end{bmatrix},
\end{align}
Here, $m$ has become the Mie order $n$, $\phi$ is the azimuthal and $\theta$ the polar angle. 
The Riccati–Bessel functions of order $n$ are denoted by $z_n$. The kind of $z_n$ is indicated by the superscript $^{(k)}$ of $\mathbf{N}$ and $\mathbf{M}$: $^{(1)}$ denotes $\psi_n$ while $^{(3)}$ denotes $\xi_n$, and their corresponding logarithmic derivatives, $D_{n}^{(1)}$ and $D_{n}^{(3)}$. These functions are given by $\psi_{n}(\rho)=\rho j_{n}(\rho)$ and $\xi_{n}(\rho)=\rho h_{n}^{(1)}(\rho)$.

\subsection{Mie coefficients}\label{app:mie_coeff}
\newcommand{\psiB}[2]{\psi_{#2}(#1)}
\newcommand{\chiB}[2]{\chi_{#2}(#1)}
\newcommand{\xiB}[2]{\xi_{#2}(#1)}
\newcommand{\psiBd}[2]{\psi_{#2}'(#1)}
\newcommand{\chiBd}[2]{\chi_{#2}'(#1)}
\newcommand{\xiBd}[2]{\xi_{#2}'(#1)}

The formula of the scattering coefficients for a spherical particle with $L$ layers first developed by Yang\cite{yangImprovedRecursiveAlgorithm2003} and later modernized by Pe\~na and Pal \cite{penaScatteringElectromagneticRadiation2009} are given by:
\begin{align}
     a_{n} =a_{n}^{L+1}=\frac{\left[H_{n}^{a}\left(m_{L} x_{L}\right) / m_{L}+n / x_{L}\right] \psi_{n}\left(x_{L}\right)-\psi_{n-1}\left(x_{L}\right)}{\left[H_{n}^{a}\left(m_{L} x_{L}\right) / m_{L}+n / x_{L}\right] \xi_{n}\left(x_{L}\right)-\xi_{n-1}\left(x_{L}\right)},
     \\ 
     b_{n} =b_{n}^{L+1}=\frac{\left[m_{L} H_{n}^{b}\left(m_{L} x_{L}\right)+n / x_{L}\right] \psi_{n}\left(x_{L}\right)-\psi_{n-1}\left(x_{L}\right)}{\left[m_{L} H_{n}^{b}\left(m_{L} x_{L}\right)+n / x_{L}\right] \xi_{n}\left(x_{L}\right)-\xi_{n-1}\left(x_{L}\right)},
\end{align}
where $m_l$ is the relative refractive index and $x_l$ is the size parameter of each layer, $l$ with $l=1,...,L$. The size parameter for each layer is given by $x_l= k r_l$ with the wavevector in the host medium $k=k_0 n_{\text{env}}$ and the radii of layer $r_l$. $\psi_{n-1}(x_L)$ and $\xi_{n-1}(x_L)$ are found using the reccurrence relation outlined in Section \ref{app:RiccatiBessel_recurrences}, while the composite functions $H^a_n$ and $H^b_n$ are given by:

\begin{align}
H_{n}^{a}\left(m_{1} x_{1}\right)=D_{n}^{(1)}\left(m_{1} x_{1}\right),
\end{align}
\begin{align}
H_{n}^{a}\left(m_{l} x_{l}\right)=\frac{G_{2} D_{n}^{(1)}\left(m_{l} x_{l}\right)-Q_{n}^{(l)} G_{1} D_{n}^{(3)}\left(m_{l} x_{l}\right)}{G_{2}-Q_{n}^{(l)} G_{1}}, \quad l=2, \ldots, L,
\end{align}
\begin{align}
H_{n}^{b}\left(m_{1} x_{1}\right)=D_{n}^{(1)}\left(m_{1} x_{1}\right),
\end{align}
\begin{align}
H_{n}^{b}\left(m_{l} x_{l}\right)=\frac{\widetilde{G}_{2} D_{n}^{(1)}\left(m_{l} x_{l}\right)-Q_{n}^{(l)} \widetilde{G}_{1} D_{n}^{(3)}\left(m_{l} x_{l}\right)}{\widetilde{G}_{2}-Q_{n}^{(l)} \widetilde{G}_{1}}, \quad l=2, \ldots, L,
\end{align}
where,
\begin{align}
G_{1}=m_{l} H_{n}^{a}\left(m_{l-1} x_{l-1}\right)-m_{l-1} D_{n}^{(1)}\left(m_{l} x_{l-1}\right), \\
G_{2}=m_{l} H_{n}^{a}\left(m_{l-1} x_{l-1}\right)-m_{l-1} D_{n}^{(3)}\left(m_{l} x_{l-1}\right), \\
\widetilde{G}_{1}=m_{l-1} H_{n}^{b}\left(m_{l-1} x_{l-1}\right)-m_{l} D_{n}^{(1)}\left(m_{l} x_{l-1}\right), \\
\widetilde{G}_{2}=m_{l-1} H_{n}^{b}\left(m_{l-1} x_{l-1}\right)-m_{l} D_{n}^{(3)}\left(m_{l} x_{l-1}\right).
\end{align}
These are found recursively, details on $D_{n}^{(1)}$, $D_{n}^{(3)}$ are oulined in Section \ref{app:RiccatiBessel_recurrences}, while the ratio:
\begin{align}
Q_{n}^{(l)}=\frac{\psi_{n}\left(m_{l} x_{l-1}\right)}{\xi_{n}\left(m_{l} x_{l-1}\right)} / \frac{\psi_{n}\left(m_{l} x_{l}\right)}{\xi_{n}\left(m_{l} x_{l}\right)},
\end{align}
can be found via the upwards recccurence relation \cite{yangImprovedRecursiveAlgorithm2003}:
\begin{align}
Q_{n}^{(l)}=Q_{n-1}^{(l)}\left(\frac{x_{l-1}}{x_{l}}\right)^{2} \frac{\left[z_{2} D_{n}^{(1)}\left(z_{2}\right)+n\right]}{\left[z_{1} D_{n}^{(1)}\left(z_{1}\right)+n\right]} \frac{\left[n-z_{2} D_{n-1}^{(3)}\left(z_{2}\right)\right]}{\left[n-z_{1} D_{n-1}^{(3)}\left(z_{1}\right)\right]},
\end{align}
for $n=1, \ldots, N_{\max }$, with the initial condition:
\begin{align}
Q_{0}^{(l)}=\frac{\exp \left(-i 2 a_{1}\right)-\exp \left(-2 b_{1}\right)}{\exp \left(-i 2 a_{2}\right)-\exp \left(-2 b_{2}\right)} \times \exp \left(-2\left[b_{2}-b_{1}\right]\right), 
\end{align}
where $z_1 = m_lx_{l-1} = a_1+ib_1$ and $z_2=m_lx_l = a_2+ib_2$.

\subsection{Mie far-field observables}\label{app:mie_observables}

The scattering, absorption and extinction efficiencies can be obtained from the scattering coefficients as,
\begin{align}
    Q_{\text {ext }} &= \frac{1}{2\pi r_{s}}\frac{2 \pi}{k^{2}} \sum_{n=1}^{\infty}(2 n+1) \operatorname{Re}\left\{a_{n}+b_{n}\right\},\\
    Q_{\text{sca}} &= \frac{1}{2\pi r_{s}}\frac{2\pi}{ k^2}\sum_{n=1}^{\infty }(2n+1) \left ( \mid a_n  \mid^2  +  \mid b_n \mid^2 \right ), \\
    Q_{\text{abs}} &= Q_{\text{ext}} - Q_{\text{sca}}\, .
\end{align}
The cross sections are given by the efficiencies times the geometric cross section of the particle $\sigma_{\text{geo}} = \pi r^2$.

The angular dependent scattering i.e. the scattered irradiance per unit incident irradiance for perpendicularly and parallel polarised illumination light (with respect to the scattering plane), as well as for unpolarised light, are given by,
\begin{align}
    i_{par} = \mid S_2 \mid^{2}, && i_{per} = \mid S_1 \mid^{2}, && i_{unp} = \frac{i_{par}+i_{per}}{2}.
\end{align}
Where,
\begin{align}
    S_{1}(\theta) & = \sum_{n=1}^{n_{\max }} \frac{2 n+1}{n(n+1)}\left(a_{n} \pi_{n}(\mu)+b_{n} \tau_{n}(\mu)\right)\, , \\
    S_{2}(\theta) & = \sum_{n=1}^{n_{\max }} \frac{2 n+1}{n(n+1)}\left(a_{n} \tau_{n}(\mu)+b_{n} \pi_{n}(\mu)\right)\, ,
\end{align}
where $\mu = \cos\theta$.
The angular functions $\pi_n(\cos\theta)$ and $\tau_n(\cos\theta)$ are defined from the associated Legendre polynomials $P_n^1(\cos\theta)$.\cite{bohrenAbsorptionScatteringLight1998} These satisfy the recurrence relations
\begin{align}
    \pi_0(\mu) &= 0\, , \\
    \pi_1(\mu) &= 1\, , \\
    \pi_{n+1}(\mu) &= \frac{2n+1}{n}\mu \,\pi_n(\mu) - \frac{n+1}{n}\,\pi_{n-1}(\mu)\, ,\\
    \tau_n(\mu) &= n \mu \,\pi_n(\mu) - (n+1)\,\pi_{n-1}(\mu)\, .
\end{align}

\subsection{Near-field observables}\label{app:nearfields}

The near-fields in the $l$th region of the sphere can be found by matching the continuity conditions at each spherical interface, which yields following formulae:\replaced{\cite{ladutenkoMieCalculationElectromagnetic2017}}{\cite{bohrenAbsorptionScatteringLight1998}}
\begin{align}
\mathbf{E}_{l} =\sum_{n=1}^{\infty} E_{n}\left[c_{n}^{(l)} \mathbf{M}_{o 1 n}^{(1)}-i d_{n}^{(l)} \mathbf{N}_{e 1 n}^{(1)}+i a_{n}^{(l)} \mathbf{N}_{e 1 n}^{(3)}-b_{n}^{(l)} \mathbf{M}_{o 1 n}^{(3)}\right], \label{eq:E_l}
\\
\mathbf{H}_{l} =\frac{k_{l}}{\omega \mu} \sum_{n=1}^{\infty} E_{n}\left[d_{n}^{(l)} \mathbf{M}_{e 1 n}^{(1)}+i c_{n}^{(l)} \mathbf{N}_{o 1 n}^{(1)}-i b_{n}^{(l)} \mathbf{N}_{o 1 n}^{(3)}-a_{n}^{(l)} \mathbf{M}_{e 1 n}^{(3)}\right], \label{eq:H_l}
\end{align}
where $a_n^{(l)}$,  $b_n^{(l)}$,  $c_n^{(l)}$ and $d_n^{(l)}$ are the expansion coefficients. These have been found by Ladutenko, Konstantin, et al. \cite{ladutenkoMieCalculationElectromagnetic2017} to be,
\begin{align}
a_{n}^{(l)} =\frac{D_{n}^{(1)}\left(m_{l} x_{l}\right) T_{1}\left(m_{l+1} x_{l}\right)+T_{3}\left(m_{l+1} x_{l}\right) m_{l} / m_{l+1}}{\xi_{n}\left(m_{l} x_{l}\right) U\left(m_{l} x_{l}\right)}, \\
b_{n}^{(l)} =\frac{D_{n}^{(1)}\left(m_{l} x_{l}\right) T_{2}\left(m_{l+1} x_{l}\right) m_{l} / m_{l+1}+T_{4}\left(m_{l+1} x_{l}\right)}{\xi_{n}\left(m_{l} x_{l}\right) U\left(m_{l} x_{l}\right)}, \\
c_{n}^{(l)} =\frac{D_{n}^{(3)}\left(m_{l} x_{l}\right) T_{2}\left(m_{l+1} x_{l}\right) m_{l} / m_{l+1}+T_{4}\left(m_{l+1} x_{l}\right)}{\psi_{n}\left(m_{l} x_{l}\right) U\left(m_{l} x_{l}\right)}, \\
d_{n}^{(l)}  =\frac{D_{n}^{(3)}\left(m_{l} x_{l}\right) T_{1}\left(m_{l+1} x_{l}\right)+T_{3}\left(m_{l+1} x_{l}\right) m_{l} / m_{l+1}}{\psi_{n}\left(m_{l} x_{l}\right) U\left(m_{l} x_{l}\right)},
\end{align}
where,
\begin{align}
U(z)=D_{n}^{(1)}(z)-D_{n}^{(3)}(z), \\
T_{1}(z)=a_{n}^{(l+1)} \xi_{n}(z)-d_{n}^{(l+1)} \psi_{n}(z), \\
T_{2}(z)=b_{n}^{(l+1)} \xi_{n}(z)-c_{n}^{(l+1)} \psi_{n}(z), \\
T_{3}(z)=d_{n}^{(l+1)} D_{n}^{(1)}(z) \psi_{n}(z)-a_{n}^{(l+1)} D_{n}^{(3)}(z) \xi_{n}(z), \\
T_{4}(z)=c_{n}^{(l+1)} D_{n}^{(1)}(z) \psi_{n}(z)-b_{n}^{(l+1)} D_{n}^{(3)}(z) \xi_{n}(z).
\end{align}
The total fields in the background medium are given by $\mathbf{E}_{\text{tot}} = \mathbf{E}_i + \mathbf{E}_s$ and $\mathbf{H}_{\text{tot}} = \mathbf{H}_i + \mathbf{H}_s$, where,
\begin{align}
\mathbf{E}_{i}=\sum_{n=1}^{\infty}E_n\left(\mathbf{M}_{o 1 n}^{(1)}-i \mathbf{N}_{e 1 n}^{(1)}\right), \label{eq:E_i}
\\ 
\mathbf{H}_{i}=\frac{k_l}{\omega\mu}\sum_{n=1}^{\infty}E_n\left(\mathbf{M}_{e 1 n}^{(1)} + i \mathbf{N}_{o 1 n}^{(1)}\right), \label{eq:H_i}
\\
\mathbf{E}_{s}= \sum_{n=1}^{\infty} E_{n}\left(i a_{n} \mathbf{N}_{e 1 n}^{(3)}-b_{n} \mathbf{M}_{o 1 n}^{(3)}\right), \label{eq:E_s}
\\ 
\mathbf{H}_{s}= \frac{k_l}{\omega \mu} \sum_{n=1}^{\infty} E_{n}\left(-i b_{n} \mathbf{N}_{o 1 n}^{(3)}-a_{n} \mathbf{M}_{e 1 n}^{(3)}\right). \label{eq:H_s}
\end{align}
Physically, to ensure that the electromagnetic fields remain finite at the origin of the sphere, the corresponding expansion coefficients must vanish\cite{bohrenAbsorptionScatteringLight1998}, requiring $a_n^{(1)}= b_n^{(1)}=0$. Furthermore, it can be shown \cite{ladutenkoMieCalculationElectromagnetic2017} that, an inspection of Eqs. \eqref{eq:E_l}, \eqref{eq:H_l} and Eqs. \eqref{eq:E_i}-\eqref{eq:H_s}, reveals that $c_n^{L+1} = d_n^{L+1}=1$.

\subsection{Spherical bessel based implementation for testing}\label{app:hankel_deriv}

During development, we also implemented SciPy wrappers and PyTorch-based recurrences for standard spherical Bessel functions ($h^1_n(z)$, $j_n(z)$ and $y_n(z)$) to test a core-shell Mie formulation. Although computing Mie scattering directly with these functions leads to numerical instability for large or absorbing particles, providing these Bessel functions in PyTorch may be valuable for researchers tackling other physical systems.

Spherical Hankel functions can be obtained from spherical Bessel functions of first and second kind:
\begin{equation}
    h^1_n(z) = j_n(z) + \text{i}\,y_n(z)\, .
\end{equation}
Derivatives of spherical Bessel functions can be obtained by either of the following recurrences:
\begin{equation}\label{eq:deriv_sph_bessel}
    \begin{aligned}
        f_n^\prime(z) &= f_{n-1}(z) - \frac{n+1}{z}f_n(z)\, , \\
        f_n^\prime(z) &= - f_{n+1}(z) + \frac{n}{z}f_n(z)\, , \\
    \end{aligned}
\end{equation}
where $f_n(z)$ is any spherical bessel function. Note: In practice, to avoid negative orders, derivatives for $n>0$ are calculated using the first relation in Eqs.~\eqref{eq:deriv_sph_bessel}, while the derivative for $n=0$ is obtained from the second relation:
\begin{equation}\label{eq:bessel_deriv_zero_order}
        f_{0}^\prime(z) \;=\; -f_{1}(z) \, .
\end{equation}

To add automatic differentiation capabilities to the SciPy interface, we add custom PyTorch autodiff classes with manually implemented derivatives. To allow autograd for the first order derivatives of spherical Bessel functions that occur in the Mie coefficients, we need to implement analytic formulae for their second order derivatives.
Starting from the recurrence relation Eq.~\eqref{eq:deriv_sph_bessel} we substitute $n = n+1$ into the first equation to get,
\begin{equation}
    f_{n+1}^\prime(z) = f_{n}(z) - \frac{n+2}{z}f_{n+1}(z)
\end{equation}
Taking the derivative of the second equation,
\begin{equation}
    \frac{d^2}{dz^2}f_n(z) = - \frac{d}{dz}f_{n+1}(z) + n\frac{d}{dz} \left( \frac{f_n(z)}{z} \right),
\end{equation}
\begin{equation}
    f_n^{\prime\prime}(z) = - f_{n+1}^\prime(z) + n \left( \frac{f_n^\prime(z)z-f_n(z)}{z^2} \right).
\end{equation}
Rearrange this to,
\begin{equation}
    z^2f_n^{\prime\prime}(z) = -z^2f_{n+1}^\prime(z) + nzf_{n}^\prime - nf_n(z),
\end{equation}
and then substitute the modified first equation and the second equation to get,
\begin{align*}
    z^2f_n^{\prime\prime}(z) = -z^2\left(f_{n}(z) - \frac{n+2}{z}f_{n+1}(z)\right) + \\ nz \left( - f_{n+1}(z) + \frac{n}{z}f_n(z) \right) - nf_n(z).
\end{align*}
Rearrange this to get the equation for $f_n^{\prime\prime}(z)$,
\begin{equation}
    z^2f_n^{\prime\prime}(z) = f_n(z) \left ( -z^2 + n^2 - n\right ) + f_{n+1}(z) \left ( z(n+2) -nz \right ),
\end{equation}
\begin{equation}
    f_n^{\prime\prime}(z) = \frac{1}{z^2} \left [ (n^2 - n - z^2)f_n(z) + 2z f_{n+1}(z)     \right ].
\end{equation}

\bibliography{2025_jackson_pymiediff.bbl}
\end{document}